# USING SCRIPTING LANGUAGES TO TEACH PROGRAMMING

**Apostolos Syropoulos**   and   **Athanasios Stavrianos**

asyropoulos@yahoo.com    thstavr@gmail.com

**Abstract**

Nowadays, scripting programming languages like Python, Perl and Ruby are widely used in system programming, scientific computing, etc. Although solving a particular problem in these languages requires less time, less programming effort, and less concepts to be taught to achieve the desired goal, still they are not used as teaching tools. Therefore, the use of scripting languages as a teaching vehicle for programming course is very promising. On the other hand, GUI programming, when performed with such languages, is easy and rewarding, since one sees the result of her work immediately. Thus, we are sure that scripting languages combined with GUI toolkits will be the next big thing in computer education.

## INTRODUCTION

Nowadays, computers are everywhere—from smart phones, smart TVs, cars, etc. This means that users must have a basic understanding of computer technology to be able to use all these devices efficiently. Also, it seems that when one has a basic understanding of algorithmic problem solving, then such tasks can be performed more easily. More generally, a basic understanding of (algorithmic) problem solving can be beneficial to everybody including of course students at all levels.

When one has to teach programming, then it is necessary to first think how to teach programming and then to choose a programming notation to express her approach to programming. For example, when one is a follower of constructivism, then she might be tempted to use LOGO as this language has been designed by an advocate of constructivism. However, in certain cases one is "forced" to choose some programming language and then to adapt her teaching plan. Indeed, this happens quite often. For example, in the 1980s it was quite popular to teach programming using Pascal, whilst today one expects that introductory programming courses are taught in Java.

We think that the approaches we have just outlined are lying in the out extremes of a spectrum of choices. We do not think that if one follows Seymour Papert [11] ideas, then she must teach programming using LOGO! It is quite possible to "im-plement" Pupert's ideas using any programming language, since, for example, program decomposition is programming technique that closely resembles Papert's ideas. On the other hand, teaching classes should not be affected by hype. In particular, we choose to teach some programming language mainly because it fits best our teaching strategy and not because it is "cool" to use this language.

Unfortunately, in all disciplines there are zealots who advocate an idea, a programming language, an operating system, etc., no matter what happens in their own discipline. Thus, although LOGO has been proved a very valuable tool for teaching programming (e.g. see [4,7,8]),, we fell it is time to use modern, or at least relatively modern, programming languages to teach programming. Such languages are the so called *scripting languages*, that is, languages like Perl, Python and Ruby. These languages offer very rich libraries that can be used to do turtle graphics [12], GUI programming, etc. In particular, the creation of (simple) graphical user interfaces is quite interesting as all devices support such interfaces. Since scripting programming languages have been extensively used in the industry and they have shown their potential, we think it is high time to make the transition to scripting languages in computer education, as far it regards secondary education.

## PLAN OF THE PAPER

First we will argue on the reasons why in our opinion all pupils should learn computer programming. Then, we will describe some problems that one must overcome before attempting to teach programming in general. Next, we will explain why scripting languages are ideal tools to teach programming, in general, and GUI programming, in particular. Finally, we give the outline of course that could be used to teach computer programming using scripting languages to highschool pupils.

## WHY LEARNING HOW TO PROGRAM DOES MATTER

Back in 1997, Brain Harvey [5] wrote that

> [w]hen I wrote the first edition of this book, in 1984, it was controversial to suggest that not everyone has to learn to program. I was accused of elitism, of wanting to keep computers as a tool for the rich, while condemning poorer students to dead-end jobs. Today it's more common that I have to fight the opposite battle, trying to convince people why anyone should learn about computer programming. After all, there is all that great software out there; instead of wasting time on programming, I'm told, kids should learn to use Microsoft Word or Adobe Illustrator or Macromedia Director. At the same time, kids who've grown up with intricate and beautifully illustrated video games are frustrated by the relatively primitive results of their own first efforts at programming. A decade ago it was thrilling to be able to draw a square on a computer screen; today you can do that with two clicks of a mouse.

Unfortunately, the situation has not changed much in a number of countries, including Greece where we presently work and live. Nowadays, many people cor-rectly assume that pupils should be competent in informatics but unfortunately they equate this competence with a basic or somehow advanced knowledge of a word processor, a spreadsheet, and a presentation maker![1] Obviously, this kind of knowledge cannot be classified as knowledge in informatics but, rather, as key skills for entry level secretaries...We firmly believe that one can claim to have a basic understanding of informatics once she is able to write a simple computer program. But shy should anyone bother to learn how to program a computer?

Generally speaking, computer programming is about problem solving. Typically, a pupil who attends a programming class, learns how to algorithmically solve problems and how to express their solution into a programming language. Certainly, when, for example, one is taught physics and mathematics, then she also learns how to solve problems and since to some extend mathematics is like programming, one could argue that programming is almost useless. Of course this is an exaggeration since the similarity is superficial—in mathematics and physics we solve the problem and we are done whilst when solving a "computer" problem we have to solve the problem, to implement the solution and finally to, at least, verify that our solution does what it is supposed to do.

The implementation part of the programming task is very important because pupils learn how to express their thoughts in a specific formalism. Typically, one can "vaguely" express a solution in natural language but it seems quite difficult to do the same in a programming language. For beginners it is like trying to solve an everyday problem for people who live in Edwin A. Abbott's [1] *Flatland* by trying to thing in think in only two geometric dimensions. In different words, one must learn to express a solution using an expressive tool with quite limited expressive power. In this respect, programming is a very important exercise.

When one describes how to prepare a document in two columns using a word processor or how to create a photo collage using some image editor, then in reality she describes some sort of algorithmic procedure. Thus, many tasks that seem to have nothing in common with programming and algorithms are in fact algorithms. Therefore, pupils that have learned how to program can easily describe such tasks. Obviously, this skill can be used to precisely describe other non-algorithmic procedures like cooking.

In a nutshell, teaching programming to pupils can be really beneficial, provided of course that this is done in a systematic way. But are there any pitfalls when teaching programming?

It seems that the biggest problems when teaching programming, at least to beginners, is to explain what variables are and what is the meaning of the assignment command. One approach could be to say something that is close to what variables actually are, that is, memory cells. Nevertheless, this approach is suitable for students who are mature enough and have a basic understanding of computer technology. Thus, one is forced to use some sort of games. For instance, one could use the following simple game: a pupil (e.g., Peter) is asked by his schoolmates to perform a series of addition and in the end has to utter the sum of the numbers each schoolmate told him. Even, if his schoolmates say the numbers in a particular order, he must keep track of what their sum so far. A simple solution would be to allow Peter to have pieces of paper where he can write partial sums. Thus, when Jane tells him a number he sees what is written on the pice of paper that corresponds to Jane, makes the addition, and writes the new number on the paper. In order to avoid confusion, Peter gives to each paper a name, which, in the simplest case, can be the name of each schoolmate that plays the game. Clearly, the pieces of paper are *variables*, and the various operations Peter performs on them correspond to variable *update* and *access*. So far so good, but the real problem starts when writing down assignment commands. Let us give two examples that demonstrate the problem.

---

[1]Sadly, the Greek Ministry of Education has practically ousted computer education from upper-level secondary education, while in lower-lever secondary education teachers have to teach word-processing, spreadsheets, etc. In a way this is an oxymoron in a country whose leaders talk about innovation.

Let us start with the LOGO programming language. In LOGO variables are prefixed with either a double quotation mark or a colon, depending on their use. Thus, the symbol " is used when updating a variable with a make command, while the symbol : is used when accessing the value of a variable. This way, pupils think that there are two distinct "variables," something that is obviously confusing. In languages that follow the C tradition, one is allowed to write commands like the following one:

$$x = x + 1$$

When pupils see this command, they confuse it with mathematical equality and they shout that this is impossible! Of course, the use of special symbol like the assignment operator := that is used in languages that follow the ALGOL tradition (e.g., Pascal and Modula-2 are such languages) is far better, nevertheless, these languages are not exactly very popular today... On the other hand, it seems that one should explain that symbols are not definitely associated with particular denotations. For instance, sometimes the meaning of words depend on the context they are used. Also, the meanings of words change over time. Similarly, the equation sign may denote mathematical equality or it may be used to write down an assignment command. But it is crucial to explain that we have to agree from the beginning on the meaning of each and every symbol.

## SCRIPTING LANGUAGES

Back in 1998, John K. Ousterhout [10] wrote about a change in the way people write computer programs. In particular, he realized that people move away from system programming languages such as C or C++ to scripting languages. Indeed, today many software tools are implemented in scripting languages today. For example, many key components of the GNOME desktop are implemented in Python. In addition, Javascript and PHP, which are the languages that make our web pages dynamic, are scripting languages. Today, scripting languages are very popular and this can be proved by the great number of books devoted to teaching programming to both children and grown-ups . (e.g., see [2,3,13,14,15]). But what makes these languages so attractive?

Typically, a scripting language is one where programs are executed by an interpreter who may or may not translate on the fly the source to some byte-code for immediate execution. For example, Perl employs this scheme which implies that one can produce stand-alone executables. Although statically typed languages are considered more secure, scripting languages have dynamic type systems. In a sense they are typeless, mainly because it is easier to glue together components (e.g., when using the pipe operator of some Unix shell). In addition, scripting languages are tools for fast prototyping, which means that one can easily implement her ideas. Also, the fact that scripting languages are interpreted means that the user has access to the source code and so she can modify the source code if this is necessary. Another, interesting aspect of scripting languages is that they can be learned easier than conventional languages like C and Java. This has been nicely demonstrated by Patrick Jordan in his *Very Quick Comparison of Popular Languages for Teaching Computer Programming* [16] where he had "shown," but obviously had not proved, that the scripting language Python is easier to learn than BASIC, C, and Java. In particular, Jordan had compared how easy it is to write a simple program that inputs two numbers and prints their sum. Since this is a fairly easy program to write, let's see how a LOGO implementation of this "algorithm" compares with the corresponding implementations in three very popular scripting languages. Let's start with Ruby:

```
puts "enter the first addend"
a = gets.chomp.to_i
puts "enter the second addend"
b = gets.chomp.to_i
c= a+b
puts "their sum is",c
```

The implementation that follows is in Python:

```
print("enter the first addend")
a = int(input())
print("enter the second addend")
b = int(input())
c = a+b
print ("their sum is ",c)
```

The equivalent implementation in Perl is as follows:

```
print "Enther the first addend ";
$a = <STDIN>;
print "Enther the second addend ";
$b = <STDIN>;
```

```
$c = $a+$b;
print "their sum is $c\n";
```

Now let's see how one can implement this "algorithm" in LOGO. First of all, one should note that there is no standard input command in LOGO. So each LOGO implementation provides its own form of input command. Here is how it can be implemented in one "dialect":

```
print [enter the first addend]
make "a readWord
print [enter the second addend]
make "b readWord
make "c :a+:b
print [The sum is]
print :c
```

And here is how one can code the same "algorithm" coded in a second "dialect":

```
read [enter the first addend] "a
read [enter the second addend] "b
make "c :a+:b
print [The sum is]
print :c
```

Note that the second version is shorter than the first one. However, it is obvious that the LOGO code is by no means easier to write or to understand when compared to the implementation in the three scripting languages. One could say that the LOGO and Perl codes have the same difficulty. It seems that the Python code is the most easy to write and understand, followed by the Ruby code. Thus, this little "experiment" nicely demonstrates that programming in scripting languages is not so difficult to learn. Thus, these languages can be used for teaching basic algorithmic concepts. However, this is not the only reason why one should choose scripting languages to teach programming.

LOGO has nicely demonstrated that when programming produces some visual output then it is interesting. Of course, it is a matter of taste whether one opts to teach her pupils how to draw squares and polygons or how to create panels with buttons that mimic the behavior of common graphical user interfaces.

## GUI PROGRAMMING WITH SCRIPTING LANGUAGES

A GUI is a *Graphical User Interface* that allows users to interact with a computer program through icons, buttons, etc. Hence, the term GUI programming refers to toolkits and techniques that allow coders to create GUIs. Interestingly, GUI programming does not involve any new things about programming as a mental activity. In different words, constructing a GUI obeys the same principles and ideas that should be employed to construct a program that will sort a series of numbers entered from a console. However, one has to use a specific toolkit with its own API and set of design rules, which has to be introduced to pupils. The good news is that there are many and different toolkits. Some of them are extremely powerful but have a learning curve that is quite long, at least for high school pupils. Such toolkits are the GTK and the Qt toolkits. Others are much simpler and, thus, easier to learn and use. For example, Tk is such a toolkit. Let us see how one can create an empty window with Tkinter, that is, the Python interface to the Tk GUI toolkit. The code that follows builds a simple but empty window:

```
from tkinter import *
top = Tk()
# Code to add widgets will go here...
top.mainloop()
```

Note that the symbol # starts a comment that extends to the end of the line. Of course, one can now add components to create a "real" GUI application. The GUI that is shown in the following screen-shot has a label and a button, which when pressed shuts down the application.

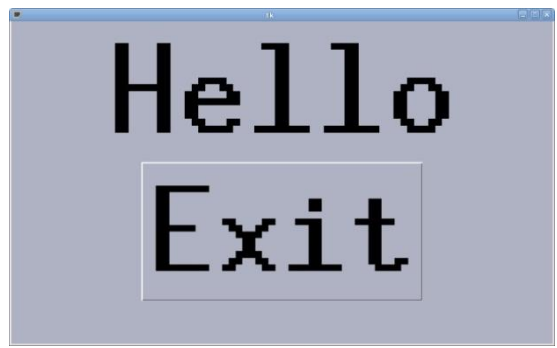

The code that follows shows exactly what should be added in order to create this GUI application:

```
from tkinter import *
# set up the window itself
top = Tk()
F = Frame(top)
F.pack()
# add the widgets
lHello = Label(F, text="Hello") #text in Greek
lHello.pack()
bQuit = Button(F, text="Terminate", command=F.quit)
bQuit.pack()
# set the loop running
top.mainloop()
```

It is not difficult to construct a GUI that plays a "game" like the one that is shown in figure 1.

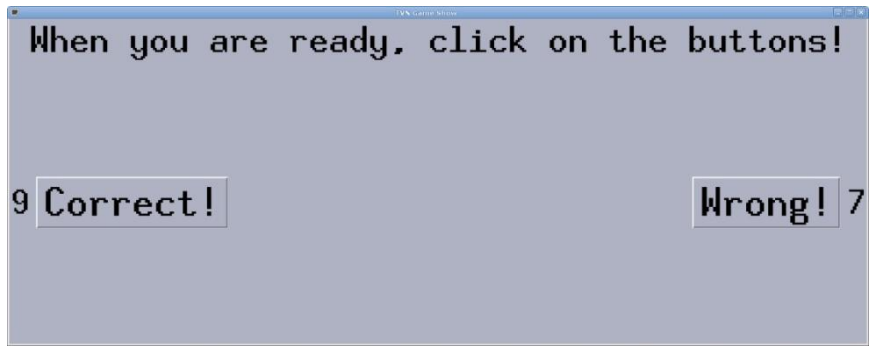

EIKÓNA 1 A GAME GUI.

In this "game," each time the user presses one of the two buttons the number next to each button is increased by one. Let's see how we can construct this GUI using Python and the Tk toolkit. First the two buttons are created by the commands that follow:

```
b1 = Button(app, text = "Correct!", width = 10,
    command = play_correct_sound)
b1.pack(side = 'left',  padx = 10, pady = 10)

b2 = Button(app, text = "Wrong!",   width = 10,
    command = play_wrong_sound)
b2.pack(side = 'right', padx = 10, pady = 10)
```

Notice that each call of the Button function includes a call to some function that is supposed to control what should happen each time a button is pressed. The code of these functions follows:

```
def play_correct_sound():
    num_good.set(num_good.get() + 1)
    correct_s.play()

def play_wrong_sound():
    num_bad.set(num_bad.get() + 1)
    wrong_s.play()
```

Of course there are some details that we have opted not to discuss, nevertheless, we do not think that these add complexity so to prohibit their use in the teaching process. For example, one such "detail" would be the association of sound clips with each GUI event. An instructor could supply to pupils her own sound clips or she could ask her pupils to create or record their own sound clips which could be used in their programs. This way, pupils would learn how to create or record audio clips for programs and how to use them.

Let us make another more realistic example. The screenshot that is shown in figure 2 is a window that asks a simple question and replies when the user chooses his answer. The window and the text of the question are rendered with the code that follows:

```
Root = Tk()
question = Label(root, text="What is the capital of Sweden?")
question.pack()
```

Each answer is a radio button of the following form:

```
R1 = Radiobutton(root, text="Athens", variable=var, value=1,
                 command=sel)
```

Here var is an integer variable wrapper that is defined as follows:

```
var = IntVar()
```

Thus, var is an object and not a simple arithmetic variable. Also, sel is a function that controls what should happen if this particular radio button is chosen. The definition of this function follows:

```
def sel() :
   if var.get() == 2:
      label.config(text = "Correct answer!")
   else:
      label.config(text = "Wrong answer!"
```

The last part of the application is easy:

```
label = Label(root)
label.pack()
Root.mainloop()
```

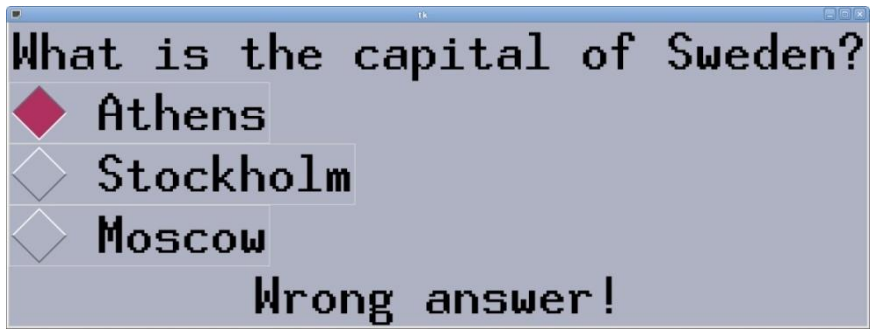

EIKÓNA 2 A SIMPLE QUIZ GUI APPLICATION.

Once pupils have mastered the GUI builder they will be able to build complete GUI applications similar to the one shown in figure 3.

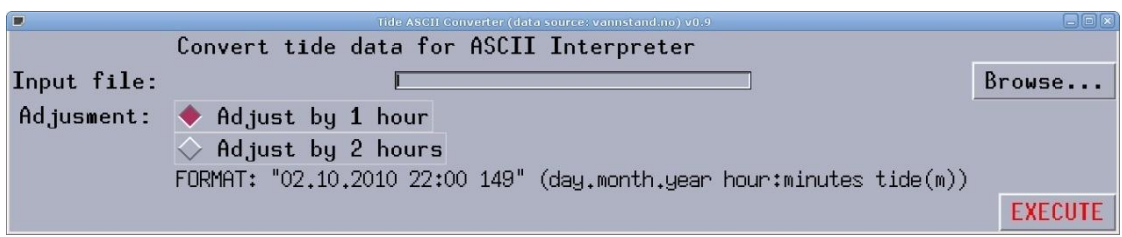

EIKÓNA 3 A COMPLETE GUI APPLICATION.

Most pupils use computers to play games, thus, it seems that many of them would be really excited to learn how to create their own games. The PyGame library is simple Python library that allows one to create simple games and other interesting things. As a demonstration of the capabilities of PyGame, we will explain how one can create the "game" shown in figure 4.

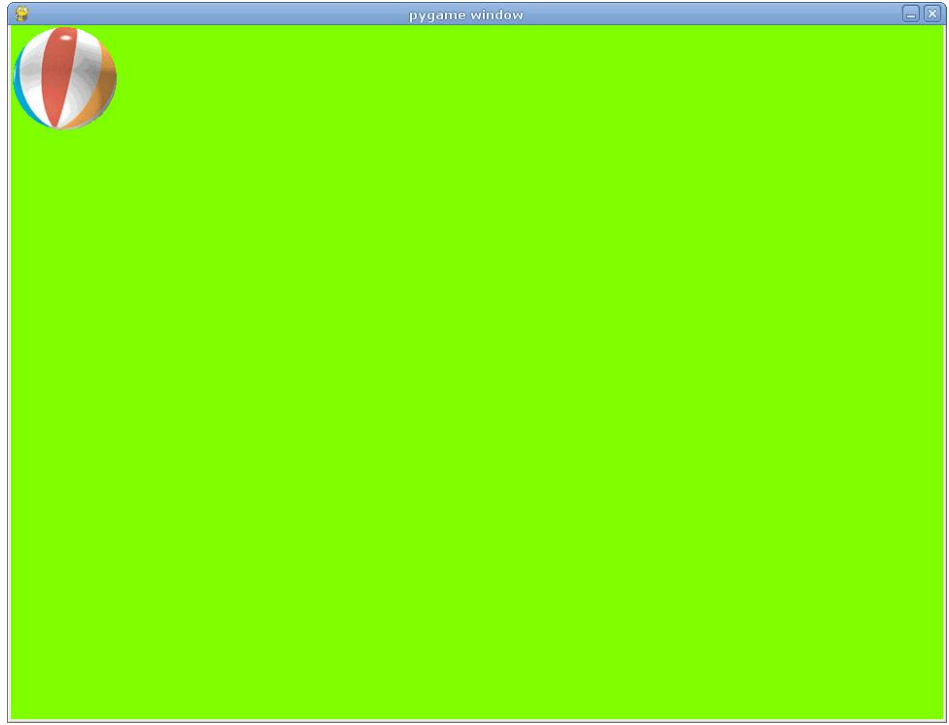

EIKÓ A SIMPLE BALL GAME. THE BALL MOVES IN THE WINDOW. WHEN IT TOUCHES AN EDGE IT BOUNCES.

First we need to initialize the window:

```
import sys, pygame
pygame.init()
size = width, height = 960, 720
screen = pygame.display.set_mode(size)
speed = [1, 1]; color = 128, 255, 0
screen = pygame.display.set_mode(size)
```

Next we need to set the ball:

```
ball = pygame.image.load("ball.gif")
ballrect = ball.get_rect()
```

The "ball" is obviously an image, thus, one could use anything for a ball. The code that follows controls the action of the ball:

```
while 1:
    for event in pygame.event.get():
    if event.type == pygame.QUIT: sys.exit()
        ballrect = ballrect.move(speed)
    if ballrect.left < 0 or ballrect.right > width:
        speed[0] = -speed[0]
```

```
            if ballrect.top < 0 or ballrect.bottom > height:
                speed[1] = -speed[1]
        screen.fill(color); screen.blit(ball, ballrect)
        pygame.display.flip()
```

Once one has introduced the various ideas and the programming tools, it would be fairly easy for pupils to make similar applications.

In general, these examples show it is easy to construct simple GUIs and relatively easy to construct more complicated GUIs. Certainly, one has to devise a syllabus that will teach pupils how to create both simple and complicated GUIs. If one wants to construct a syllabus for teaching GUI programming with Tk and some scripting language, then we feel that examples like the first one can be presented quite early to pupils. Obviously, the second example should be presented at a later stage. Naturally, one has to make sure that pupils are getting exposed to a number of basic ideas (e.g., variables, values, control structures, etc.). Of course the list of simple GUI examples is not limited to the one presented here (e.g., the interested reader can see more examples in [2]).

TABLE 1    A SYLLABUS FOR A GUI PROGRAMMING COURSE DIVIDED INTO NINE TEACHING UNITS.

| Unit Name | Unit Description |
| --- | --- |
| Introduction | Introduction to basic concepts and ideas. Brief history of Python. |
| Working with Python | Getting to know IDLE. Demonstration of simple Python applications. The `print` function. |
| Numeric Values | Variables, numeric data types, conversions, the `input` function. |
| String Values | Defining strings, using string and string operators (optionally one may discuss regular expressions). |
| Simple GUIs | Creating simple, "static" GUIs with labels and buttons only. |
| Programming Constructs | Repetition and conditional commands. |
| Functions | The notion of function. Defining new functions. Binding events to widgets. |
| Data Structures | Introducing hash tables and/or arrays. |
| Advanced GUIs | Check boxes, radio buttons, list boxes, etc. |

## A SYLLABUS FOR GUI PROGRAMMING

In Table 1 we present the outline of a syllabus that can be used as a basis for teaching GUI programming with Python and the Tk toolkit. This syllabus is based on the assumption that programming is taught two times each week for 50 minutes. One can use Ruby, Perl, or Lua instead of Python without extensive modifications to this syllabus. In fact, the required changes are very simple. For example, in the case of Perl one needs to make sure that pupils understand that the dollar sign is part of a variables name. Also, if opting to use Ruby, then users should use *Shoes* which is an alternative toolkit that is extremely simple and powerful. Furthermore, since this syllabus is an outline it can be easily adapted to suit the special needs of a particular school.

A few years ago, it was possible to choose any programming language to teach programming in Greek schools. Unfortunately, this is not valid anymore and so we cannot apply our ideas and present experimental results. However, when it was possible to chose the programming language to use in teaching classes, one of us had used Perl. Although, it was not possible to teach GUI programming, still pupils had no difficulty understanding programming in Perl. Thus, it seems quite reasonable to expect that students will have no difficulty learning GUI programming with scripting languages.

## CONCLUSIONS

We have explained why in our opinion, programming should be taught in secondary education. Also, we have presented scripting languages and have shown why in our opinion these languages can be easily used to teach programming to high school pupils. Also, we have explained why GUI programming can be particularly appealing to pupils since it is not difficult to construct simple and relatively complicated GUIs using any scripting language. Moreover, we have presented a syllabus that can be used to GUI programming with a scripting language.

In general, all popular scripting languages are open source systems. These languages can be taught using proprietary systems, but a more natural choice would be to teach them using an open source systems. This way, pupils will be exposed to open source systems (e.g., GNU/Linux, FreeBSD, OpenBSD, OpenIndiana, etc.) which, in our opinion, is very important. For example, pupils would learn the benefits of collaboration, especially between people who do not live and work in the same physical location. Thus, pupils would be taught that the Internet can be used a productive medium and it is not only for chatting, posting photos, and downloading files. A side effect would be the creation of a trend in using and working with open source systems.

We think that pupils that have been introduced to programming with some scripting language at lower high school, they can easily move to some more "advanced" programming language to upper high school. For example, Scala [9] would be an ideal "advanced" language, particularly because it is possible to write Scala scripts. This way, pupils will be forced to initially learn a different syntax but not a different execution model. Naturally, later on they can be instructed how to compile their source code. Of course, this later part of Scala should be available only to pupils that may like to pursue higher education in fields related to computer programming.